\documentclass[prl,twocolumn, aps, amsmath, amssymb,superscriptaddress, footinbib]{revtex4-1}
\usepackage{graphicx,amsmath,amssymb}
\usepackage[usenames]{color}

\def\beq{\begin{equation}}
\def\eeq{\end{equation}}
\def\be{\begin{eqnarray}}
\def\ee{\end{eqnarray}}
\def\beq{\begin{equation}}
\def\eeq{\end{equation}}
\def\bea{\begin{eqnarray}}
\def\eea{\end{eqnarray}}

\def\Beq{\begin{equation}\begin{aligned}}
\def\Eeq{\end{aligned}\end{equation}}

\DeclareRobustCommand{\SkipTocEntry}[4]{}

\newcommand{\reading}[1]{\hfill $ _{\text{\hyperref[#1]{ref}}} $}

\begin{document}
\title{On large $q$ expansion in the Sachdev-Ye-Kitaev model}
\author{Grigory Tarnopolsky}
\affiliation{Department of Physics, Harvard University, Cambridge MA
02138, USA} 
\begin{abstract}
We consider the Sachdev-Ye-Kitaev (SYK) model where interaction involves $q$ fermions at a time. We find the next order correction to the thermal two-point function in the large $q$ expansion. Using this result we find the next order correction to the SYK free energy. 
\end{abstract}
\maketitle
\section{Introduction}
The Sachdev-Ye-Kitaev (SYK) model is a quantum mechanical model of $N$  interacting Majorana fermions $\chi_{i}$,  $i=1,\dots, N$ with the Hamiltonian \cite{Sachdev:1992fk,Kitaev:2015}:
\begin{align}
H_{\textrm{SYK}_{q}}= (i)^{\frac{q}{2}}\sum_{1\leq i_{1}<i_{2}<\dots<  i_{q}\leq N}J_{i_{1}\dots i_{q}}\chi_{i_{1}}\chi_{i_{2}}\dots \chi_{i_{q}}\,,
\end{align}
where $\{\chi_{i},\chi_{j}\}=\delta_{ij}$ and $J_{i_{1}\dots i_{q}}$ are  random couplings drawn from a Gaussian distribution
with zero mean and a width $\langle J_{i_{1}\dots i_{q}}^{2}\rangle = (q-1)! J^2/N^{q-1}$. 
One is usually interested in computing correlation functions, and  particularly 
 two-point  function at temperature $T=1/\beta$:
\begin{align}
G(\tau) = \langle T\chi (\tau)\chi (0)\rangle_{\beta}\,.
\end{align}
At the large $N$ limit only  melonic Feynman diagrams contribute
to the two-point function in the SYK model. These diagrams can be resummed 
and one obtains a non-perturbative Schwinger-Dyson equation: 
\begin{align}
G(i\omega_{n})^{-1}=-i\omega_{n}-\Sigma(i\omega_{n}), \quad \Sigma(\tau) = J^{2}G(\tau)^{q-1}\,, \label{SD}
\end{align}
where $G(i\omega_{n}) =\int_{0}^{\beta}d\tau e^{i\omega_{n}\tau}G(\tau)$ and $\omega_{n}=2\pi \beta^{-1}(n+1/2)$.
It is not possible to solve this equation analytically, but one can find  solution in the infrared limit, where 
$\omega$ is small and the bare $-i\omega_{n}$-term in (\ref{SD}) can be neglected \cite{Sachdev:1992fk,Kitaev:2015, Maldacena:2016hyu, PhysRevB.59.5341}: 
\begin{align}
G_{c}(\tau)=b\Big(\frac{\pi}{\beta \sin \frac{\pi \tau}{\beta}}\Big)^{2/q} \textrm{sgn}(\tau)\,, 
\end{align}
where $J^{2}b^{q}\pi=(1/2-1/q)\tan(\pi/q)$. Nevertheless it is still interesting to obtain some analytic approximation for 
$G(\tau)$ which interpolates both UV and IR regions. One way to proceed is to use the large $q$ expansion. 
The first order in $1/q$ was found in \cite{Maldacena:2016hyu}. In this note we compute the next $1/q^2$ correction 
and argue that it improves the approximation significantly, such that it  agrees with numerical results quite well.

At the next section we compute $1/q^2$ correction to the two-point function. Next we compare 
the large $q$ results and numerics. At the end we compute the large $q$  free energy and 
the coefficient of the Schwarzian action.

\section{Large $q$ two-point function}
We consider the large $q$ ansatz for the two-point function \cite{Maldacena:2016hyu}:
\begin{align}
G(\tau) = \frac{1}{2}\textrm{sgn}(\tau)\Big(1+\frac{1}{q}g(\tau)+\frac{1}{q^{2}}h(\tau)+\dots\Big)\,. \label{Gtau}
\end{align}
For the self-energy (\ref{SD}) we find (we assume that $q$ is even)
\begin{align}
\Sigma(\tau)
&=\frac{\mathcal{J}^{2}}{q}\textrm{sgn}(\tau) e^{g}\Big(1+\frac{1}{q} \big(h-g-\frac{1}{2}g^{2}\big)+\dots\Big)\,, 
\label{selfenerg} 
\end{align}
where  a new coupling constant 
$\mathcal{J}^{2}= 2^{1-q}q J^{2}$ is introduced. From now on we work on the interval $\tau\in[0,\beta]$ and we can omit $\textrm{sgn}(\tau)$ in all formulas. Expanding $G(i\omega_{n})^{-1}$ in $1/q$ series up to $1/q^{2}$ term  using (\ref{Gtau}) we obtain   
\begin{align}
G(i\omega_{n})^{-1}  =&-i\omega_{n} +\frac{1}{2q}\omega_{n}^{2}g(i\omega_{n})\notag\\
&+\frac{\omega_{n}^{2}}{2q^{2}}\Big(h(i\omega_{n})+\frac{i\omega_{n}}{2}g*g(i\omega_{n})\Big)\,,
\end{align}
where  $g*g(i\omega_{n}) \equiv \int_{0}^{\beta}d\tau e^{i\omega_{n}\tau}g^{2}(\tau)$. Then using the equations (\ref{SD}) and (\ref{selfenerg}) and going back to the coordinate space we find  differential equations for each order of $1/q$:
\begin{align}
&\partial_{\tau}^{2}g =2\mathcal{J}^{2}e^{g}\,,\notag\\
&\partial_{\tau}^{2}h =2\mathcal{J}^{2}e^{g} h +\frac{1}{2} \partial_{\tau}^{3}(g*g) -2\mathcal{J}^{2}e^{g}\big(g+\frac{1}{2}g^{2}\big)\,,
\end{align}
and the functions $g(\tau)$ and $h(\tau)$ satisfy the boundary conditions $g(0)=g(\beta)=0$ and $h(0)=h(\beta)=0$.
Now we introduce a convenient variable $x= \frac{\pi v}{2}-\frac{\pi v \tau}{\beta}$. Then the first equation has the solution 
\begin{align}
g(x) = \log \Big(\frac{\cos \frac{\pi v}{2}}{\cos x}\Big)^{2}, \quad \beta \mathcal{J} =\frac{\pi v}{\cos\frac{\pi v}{2}}\,. \label{gsol}
\end{align}
Using this solution the second equation can be represented as 
\begin{align}
&\Big(\partial_{x}^{2}-\frac{2}{\cos^{2} x}\Big)h(x) =\notag\\
&\quad\quad = -\frac{\pi v}{\beta}\frac{1}{2}\partial_{x}^{3}(g*g) -\partial_{x}^{2}g(x)\big(g(x)+\frac{1}{2}g^{2}(x)\big)\,.
\end{align}
The solution to this  equation can be written as
\begin{align}
h(x) =& -\int_{-\frac{\pi v}{2}}^{\frac{\pi v}{2}} dy \mathcal{G}(x,y) \Big(\frac{\pi v}{\beta}\frac{1}{2}\partial_{y}^{3}(g*g) \notag\\
&\qquad\qquad+\partial_{y}^{2}g(y)\big(g(y)+\frac{1}{2}g^{2}(y)\big)\Big)\,, \label{answh1}
\end{align}
where the  Green's function $\mathcal{G}(x,y)$ obeys  the equation 
\begin{align}
\Big(\partial_{x}^{2}-\frac{2}{\cos^{2} x}\Big)\mathcal{G}(x,y)=\delta(x-y)\,
\end{align}
with the boundary conditions $\mathcal{G}(-\frac{\pi v}{2},y)=\mathcal{G}(\frac{\pi v}{2},y)=0$.
One can  solve this equation and obtain an explicit formula for the Green's function
\begin{align}
&\mathcal{G}(x,y) =\frac{1}{2 V}(\tan x_{<}(V+x_{<})+1)(\tan x_{>}(V-x_{>})-1)\,, \label{Green}
\end{align}
where $V\equiv \frac{\pi v}{2}+\cot\frac{\pi v}{2}$ and $x_{>}\equiv\max(x,y)$ and $x_{<}\equiv\min(x,y)$. 
Computing the convolution  
\begin{align}
&\frac{\pi v}{\beta}\frac{1}{2}\partial_{x}^{3}(g*g) =\notag\\
& =2\partial_{x}\Big(g(x)\big(\cot(\frac{\pi v}{2}+x)-\cot(\frac{\pi v}{2}-x)\big)\Big)-4\,,
\end{align}
and using the explicit formula for the Green's function (\ref{Green}) we  obtain from (\ref{answh1}) 
\begin{align}
&h(x) =\frac{1}{2}g^{2}(x)-2\ell(x) -4  \Big(\tan x\int_{0}^{x} dy \ell(y) +1\Big) \notag\\
&\quad +4 \frac{1+x\tan x}{1+\frac{\pi v}{2}\tan\frac{\pi v}{2}} \Big(\tan \frac{\pi v}{2}\int_{0}^{\frac{\pi v}{2}} dy \ell(y) +1\Big)\,, \label{hans}
\end{align}
where $\ell(x) \equiv g(x) -e^{-g(x)}\textrm{Li}_{2}(1-e^{g(x)})$ and $g(x)$ is given in (\ref{gsol}). One can compute explicitly the integral (formulas from \cite{Davydychev:2003mv} are useful)
\begin{align}
\int_{0}^{\frac{\pi v}{2}} dy \ell(y) = -\frac{\pi ^2 v^2 }{24 \cos ^2\frac{\pi  v}{2}}(\pi  v+3 \sin \pi  v)\,. \label{mint}
\end{align}

\section{Comparison with numerical results}
In this section we compare the large $q$ result with the numerical solution of the Schwinger-Dyson equation (\ref{SD}). In general we expect the large $q$ formula to work well when $|g(\tau)|\ll q$ and $|h(\tau)|\ll q^2$. These inequalities are fulfilled when $\beta \mathcal{J} \ll \pi e^{q/2}$. 

\begin{figure}[h!]
\centering
\begin{tabular}{cc}
\includegraphics[width=0.53\textwidth]{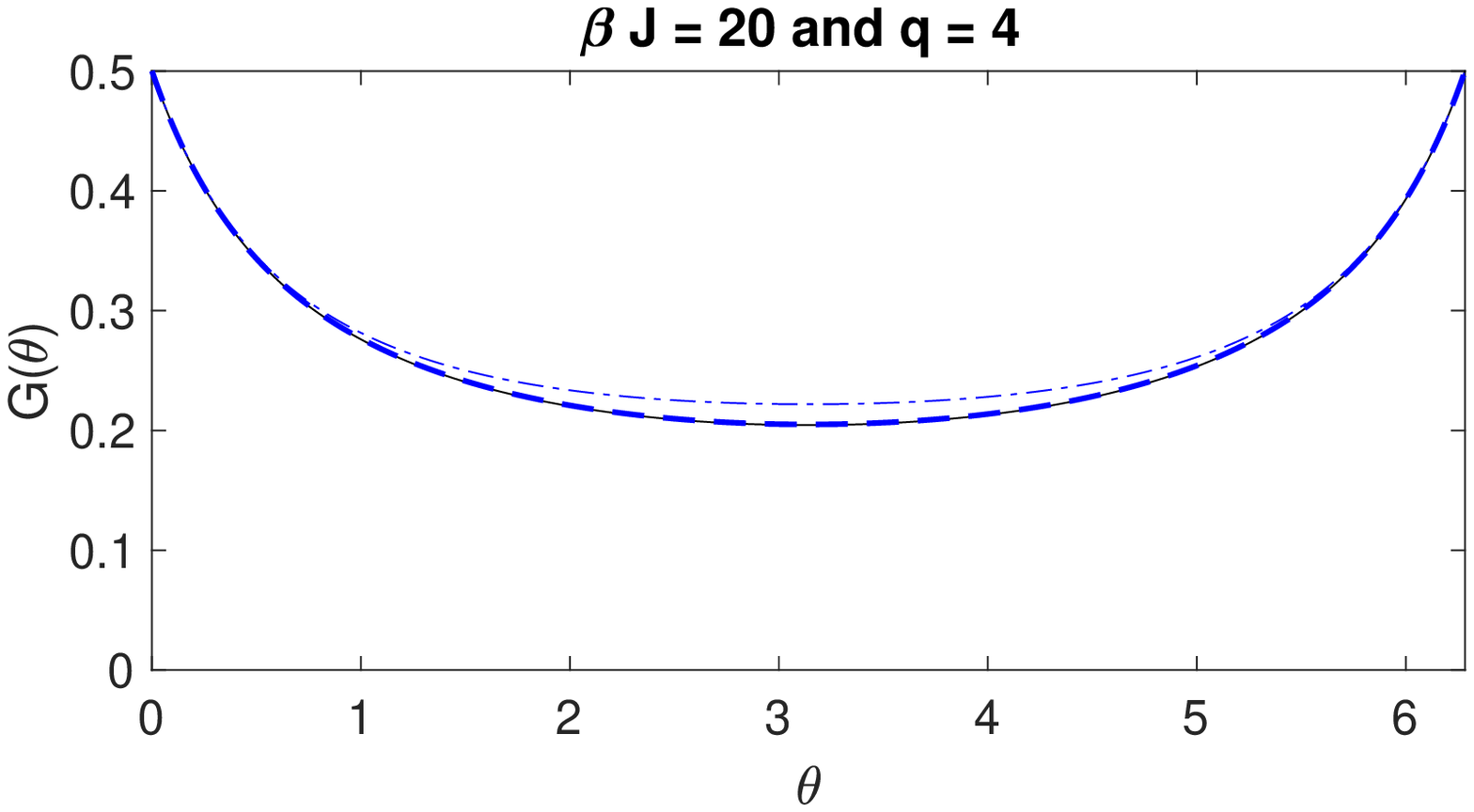}\\
\includegraphics[width=0.53\textwidth]{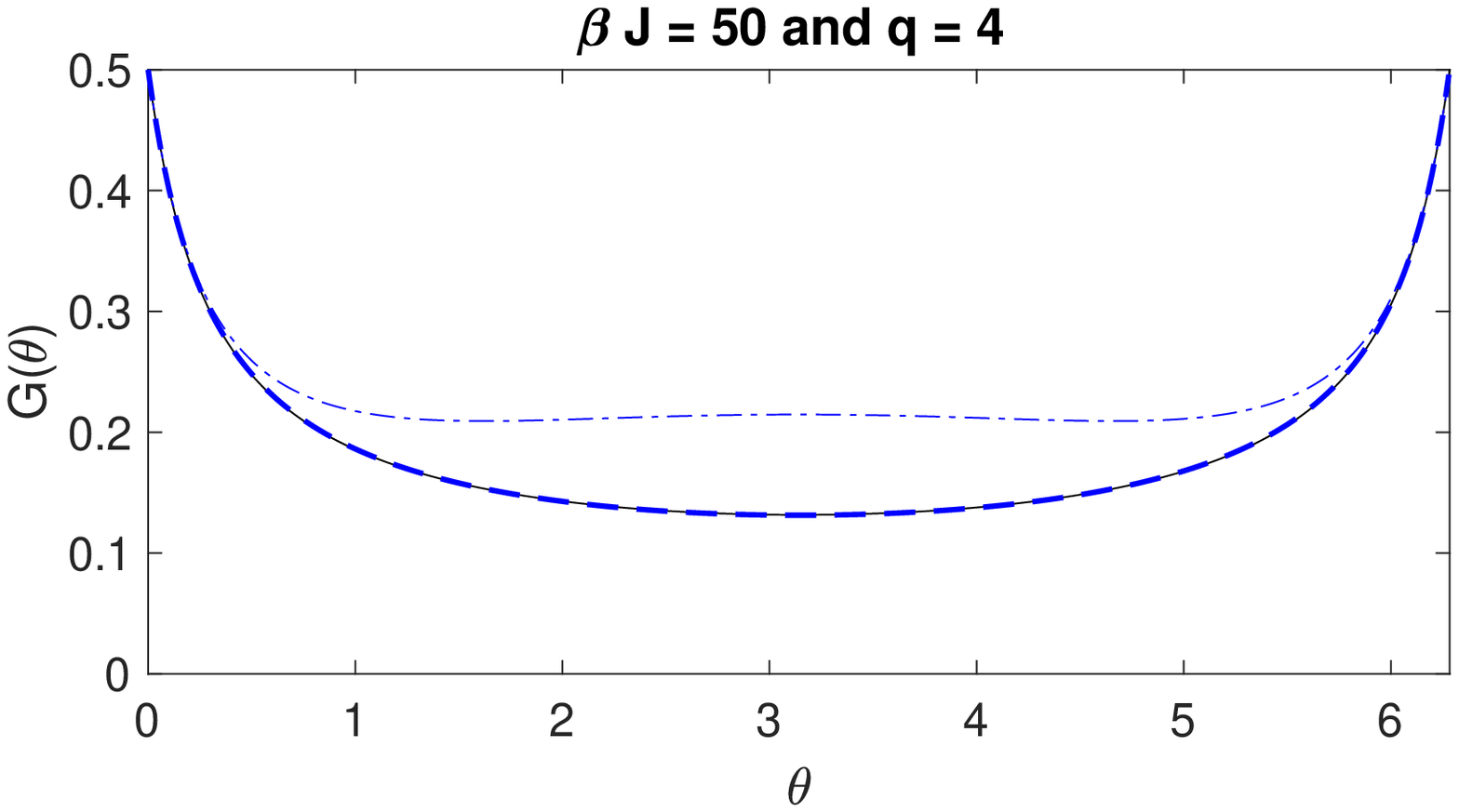}\\
\includegraphics[width=0.53\textwidth]{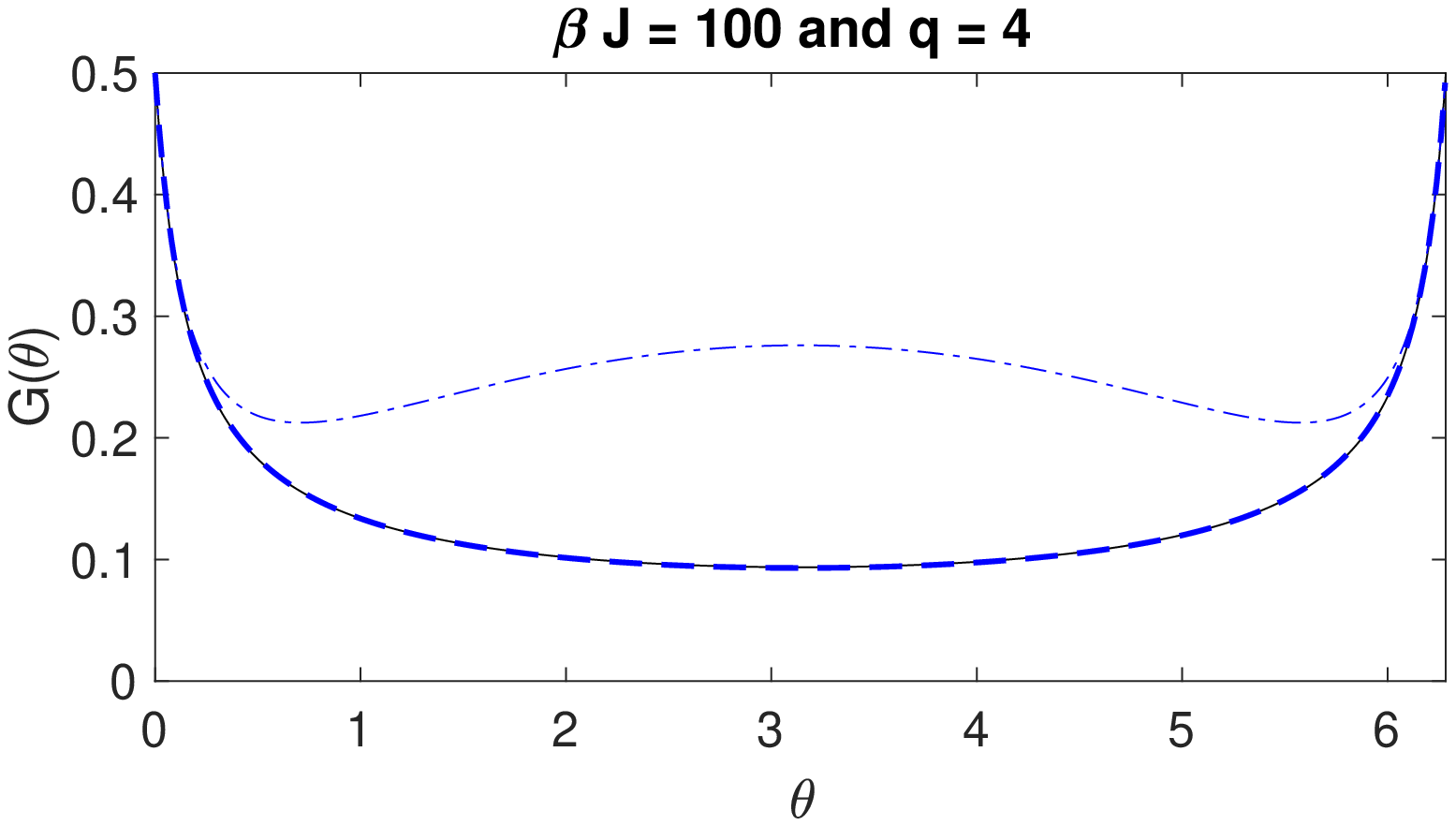}\\
\includegraphics[width=0.53\textwidth]{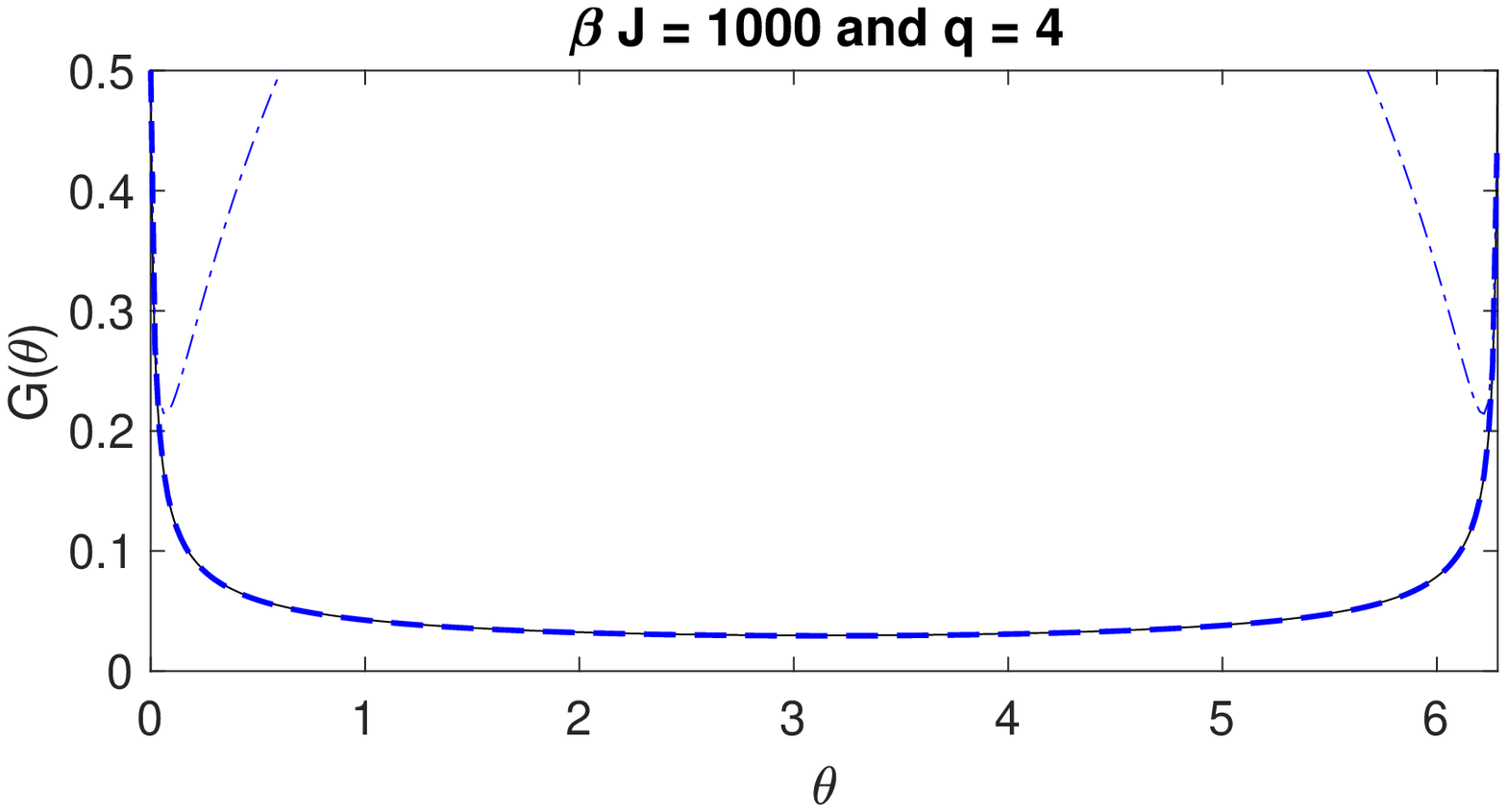}
\end{tabular}
\caption{\label{2ptfun} (Color online) Plots of  the numerical solution  and the large $q$ approximations for $G(\theta)$, $\theta=2\pi \tau/\beta$ at $\beta J = 20,50,100,1000$ and $q=4$. The black solid line is the numerical solution for the Schwinger-Dyson equation (\ref{SD}). The blue dash-dotted line is the large $q$ approximation (\ref{Gtau}) with $1/q^2$ term. The blue dashed  line is the exponentiated two-point function  (\ref{Gimpr}). }
\end{figure}

Looking at the explicit formula (\ref{hans}) it is tempting to exponentiate the result and to introduce an exponentiated  large $q$ two-point function 
\begin{align}
G(\tau)= \frac{1}{2}\textrm{sgn}(\tau)\exp\Big(\frac{1}{q}g+\frac{1}{q^{2}}(h-\frac{1}{2}g^{2})\Big)\,, \label{Gimpr}
\end{align}
which is equivalent to (\ref{Gtau}) up to order $1/q^2$.
We plot numerical and the large $q$ results for $q=4$ and different values of $\beta J$ in figure \ref{2ptfun}. We can see that the exponentiated result 
works very precisely even for large $\beta J$, whereas the large $q$ answer  (\ref{Gtau})   deviates significantly from numerics at large $\beta J$. 

\section{Large $q$ free energy}
The leading large $N$ approximation to the free energy in the SYK model is \cite{Kitaev:2015, Sachdev:2015efa}
\begin{align}
&-\frac{\beta F}{N} = \log \textrm{Pf}(\partial_{\tau}-\Sigma)\notag\\
&\qquad-\frac{1}{2}\int_{0}^{\beta} d\tau_{1}d\tau_{2}\Big(\Sigma(\tau_{12})G(\tau_{12})-\frac{J^{2}}{q}G(\tau_{12})^{q}\Big)\,.
\end{align}
To avoid evaluating the Pfaffian it is convenient to differentiate the free energy by $J\partial_{J}$ \cite{Maldacena:2016hyu}
\begin{align}
J\partial_{J}(-\beta F/N)& = -\frac{\beta}{q}\partial_{\tau}G|_{\tau\to +0}\notag\\
&=\frac{\pi v}{2q}\big(\frac{1}{q}\partial_{x}g+\frac{1}{q^{2}}\partial_{x}h\big)|_{x\to \frac{\pi v}{2}}\,,
\end{align}
where from (\ref{gsol}) and (\ref{hans}) we find
\begin{align}
&\partial_{x}g|_{x\to \frac{\pi v}{2}} = 2 \tan\frac{\pi v}{2}\,, \notag\\
&\partial_{x}h|_{x\to \frac{\pi v}{2}} = \frac{4}{1+\frac{\pi v}{2}\tan \frac{\pi v}{2}}\Big(\frac{\pi v}{2}\notag\\
&\quad\qquad-\tan \frac{\pi v}{2}(1+\frac{\pi v}{2}\tan \frac{\pi v}{2})-\int_{0}^{\frac{\pi v}{2}}dy \ell(y)\Big)\,.
\end{align}
Next, using (\ref{mint}) and   
\begin{align}
J\partial_{J}= \frac{v\partial_{v}}{1+\frac{\pi v}{2}\tan \frac{\pi v}{2}}\,,
\end{align}
we can integrate back and  obtain  $-\beta F/N = \frac{1}{2}\log 2 +\frac{1}{q^{2}}F_{1/q^{2}}+ \frac{1}{q^{3}}F_{1/q^{3}}+\dots$, where
\begin{align}
&F_{1/q^{2}}(v)= \pi v \Big(\tan \frac{\pi v}{2}- \frac{\pi v}{4}\Big)\,,\notag\\ 
& F_{1/q^{3}}(v) =\pi  v \Big(\pi  v-2  \tan \frac{\pi  v}{2} \big(1-\frac{\pi ^2 v^2}{12}\big)\Big)\,. \label{F2ordres}
 \end{align}
Expanding the free energy at strong coupling  by using that 
\begin{align}
v = 1 -\frac{2}{\beta \mathcal{J}} +\frac{4}{(\beta \mathcal{J})^{2}} -\frac{(24+\pi^{2})}{3 (\beta \mathcal{J})^{3}}+\dots\,,
 \end{align}
we find 
\begin{align}
-\frac{\beta F}{N} =&\beta \mathcal{J}\Big(\frac{1}{q^{2}}-\frac{12-\pi^{2}}{6q^{3}}\Big)+\Big(\frac{1}{2}\log 2 -\frac{\pi^{2}}{4q^{2}}+\frac{\pi^{2}}{3q^{3}} \Big)\notag\\
&+
\frac{1}{\beta \mathcal{J}}\Big(\frac{\pi^{2}}{2q^{2}}-\frac{\pi^{2}(\pi^{2}+12)}{12 q^{3}}\Big)\notag\\
&+
\frac{1}{(\beta \mathcal{J})^{2}}\Big(-\frac{\pi^{2}}{q^{2}}+\frac{\pi^{2}(5\pi^{2}+24)}{9 q^{3}}\Big)+\dots, \label{Freeendec}
 \end{align}
where the first three terms are   the ground state energy,  the zero-temperature entropy and the  temperature dependent correction to the entropy.
The zero temperature entropy coincides with the large $q$ expansion of the formula \cite{Kitaev:2015, PhysRevB.63.134406}
\begin{align}
\frac{S_{0}}{N} &= \frac{1}{2}\log 2 -\int_{0}^{1/q} d x \pi (\frac{1}{2}-x) \tan \pi x \,. 
\end{align}
The last term in (\ref{Freeendec}) agrees with the formula reported in \cite{Cotler:2016fpe, Jevicki:2016ito, Kitaev:2017awl}.

\begin{figure}[h!]
\centering
\begin{tabular}{cc}
\includegraphics[width=0.48\textwidth]{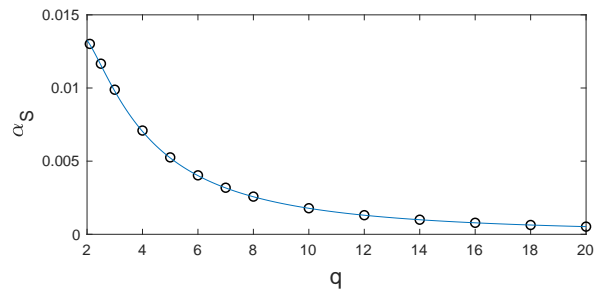}
\end{tabular}
\caption{\label{aspade} (Color online) Plot of $\alpha_{S}$ as a function of $q$. The black circles 
correspond to numerical results adapted from \cite{Maldacena:2016hyu}. The blue solid line corresponds to the two-sided Pade approximation (\ref{padeas}).  }
\end{figure}

Using the result (\ref{Freeendec}) one can find the coefficient of the Schwarzian action. The Schwarzian action, which governs the low energy dynamics of the SYK model is given by the formula \cite{Maldacena:2016hyu, Maldacena:2016upp, Engelsoy:2016xyb, Jensen:2016pah, Jevicki:2016bwu}
\begin{align}
S = -N \frac{\alpha_{S}}{\mathcal{J}} \int d\tau \{f,\tau\} , \quad  \{f,\tau\}\equiv \frac{f'''}{f'}-\frac{3}{2}\Big(\frac{f''}{f'}\Big)^{2}\,,
\end{align}
where the coefficient $\alpha_{S}$ depends on $q$. This  coefficient is related to 
the finite temperature correction to the free energy, so at large $q$ using (\ref{Freeendec}) we find 
\begin{align}
-\frac{\beta F}{N} \supset \frac{2\pi^{2}\alpha_{S}}{\beta \mathcal{J}} = \frac{1}{\beta \mathcal{J}}\Big(\frac{\pi^{2}}{2q^{2}}-\frac{\pi^{2}(\pi^{2}+12)}{12 q^{3}}+\dots\Big)\,.
\end{align} 
 At $q=2$ one has $\alpha_{S}=\frac{1}{24\pi}$.
Using asymptotics for $\alpha_{S}$
at $q=\infty$ and $q=2$
\begin{align}
\alpha_{S}(q) =\begin{cases}
\frac{1}{24\pi}+\dots\,,\;\;\qquad\qquad q\to 2\\
\frac{1}{4q^{2}}-\frac{\pi^{2}+12}{24q^{3}}+\dots, \quad  q\to \infty
\end{cases}
\end{align}
we  obtain two-sided Pade approximant:
\begin{align}
\textrm{Pade}_{[3,1]}: \;\; \alpha_{S}(q) =\frac{\pi ^2-18 \pi +24+3 (3 \pi -2) q}{6 q^2 \left(\pi ^3+8+2 (3 \pi -2) q\right)}\,.\label{padeas}
\end{align}
We note that one can improve approximation by using more terms near $q=2$ \cite{Jevicki:2016ito}.
We plotted Pade approximation and numerical results adapted from \cite{Maldacena:2016hyu} in figure \ref{aspade}.
We see that the Pade approximation is very close to numerics.

\section{Conclusions}
It would be interesting to generalize the result of this article to other SYK-type models,
discussed in \cite{Gross:2016kjj, Davison:2016ngz, Gu:2016oyy, Fu:2016vas}. Especially  it would be interesting to compute the thermalization time using large $q$ solution 
for the SYK models discussed in \cite{Eberlein:2017wah}.

It is also interesting to develop $1/q$ expansion for the higher dimensional SYK models \cite{Berkooz:2016cvq,Turiaci:2017zwd, Murugan:2017eto,Giombi:2017dtl, Prakash:2017hwq} where the stability 
of the large $N$ limit is unclear.

The large $q$ approximation to the two-point function can be  used as well in studying tensor models \cite{Witten:2016iux, Gurau:2016lzk, Klebanov:2016xxf}. Even though the general $q$ melonic tensor interaction have some  ambiguities \cite{CJepsen, Klebanov:2016xxf}, one can just formally consider large $q$ generalization of the Schwinger-Dyson equation.

\vskip 5 pt

G.T. would like to thank Yingfei Gu, Igor Klebanov, Subir Sachdev and Douglas Stanford for useful comments.  Also G.T. thanks Douglas Stanford  for providing numerical results for $\alpha_{S}$ from  \cite{Maldacena:2016hyu}.  This research was  supported  by the MURI grant W911NF-14-1-0003 from ARO and by DOE grant de-sc0007870.







\bibliography{SYKlargeq}

\end{document}